\documentclass{moriond}

\def\be{\begin{equation}}
\def\ee{\end{equation}}
\def\bea{\begin{eqnarray}}
\def\eea{\end{eqnarray}}

\newcommand{\Photo}{}
\usepackage{graphicx}
\usepackage{subfigure}

\begin{document}

\begin{flushleft} 
\mbox{CERN-PH-TH-2014-095}
\end{flushleft}

\vspace*{4cm}
%\title{The soft-virtual Higgs cross-section at N3LO}
\title{The Soft-Virtual Higgs Cross-section at N3LO and the Convergence of the Threshold Expansion}

\author{ FRANZ HERZOG$^1$ and BERNHARD MISTLBERGER$^2$}

\address{${}$\\ $^1$CERN Theory Division, CH-1211, Geneva 23, Switzerland \\$^2$Institute for Theoretical Physics, ETH Zurich, 8093 Zurich, Switzerland}

\maketitle\abstracts{We discuss the validity of the soft-virtual approximation and the threshold expansion for the Higgs boson production cross-section at hadron colliders in perturbative QCD up to next-to-next-to-next-to-leading order (N${}^3$LO). }

\section{Introduction}
Studying the properties of the Higgs boson, which was recently discovered by the ATLAS and CMS collaboration at the LHC  \cite{HiggsDiscovery}, demands high precision prediction for experimental results. Furthermore, to be able to distinguish Standard Model (SM) physics from possible deviations a precise theoretical knowledge of the predictions for the experimental outcome is vital.
% As the amount of collected data increases determining the strengths of the interactions of the Higgs boson will soon be limited by insufficiently precise perturbative quantum field theoretical predictions.
Soon the determination of the strengths of the Higgs bosons interactions will be limited by insufficiently precise 
%perturbative quantum field theoretical% 
predictions.

The Higgs production cross-section at the LHC takes the form
\begin{equation}
\label{eq:sigma}
\sigma = \sum_{ij} \int dx_1\, dx_2\, f_i(x_1)\,f_j(x_2)\, \hat{\sigma}_{ij}(m_H^2,x_1\,x_2\,s)\,,
\end{equation} 
where  $\hat{\sigma}_{ij}$ are the partonic cross-sections for producing a Higgs boson from partons $i$ and $j$, $f_i(x_1)$ and $f_j(x_2)$ are the corresponding parton distribution functions, and $m_H$ and $s$ denote the mass of the Higgs boson and the hadronic centre-of-mass energy, respectively. 
The largest contribution to the partonic cross-section is given by the gluon-fusion production mode creating a Higgs boson via a top quark loop that is formed via the interaction of two initial state gluons. 
The relatively light mass of the Higgs boson allows for the calculation of this process in perturbative QCD in the infinite top-quark mass limit.
Currently, the gluon-fusion cross-section is known in this fixed order QCD approximation up to next-to-next-to-leading order (NNLO) \cite{nnlo2,nnlo} and many additional corrections are available (see ref.\cite{ihixs} for a comprehensive summary).

The largest perturbative uncertainty of the partonic cross-section originates from the missing next-to-next-to-next-to-leading order (N${}^3$LO) QCD corrections to the gluon fusion production channel. Recently the first term of a threshold expansion, of the N${}^3$LO gluon fusion channel, was made public in the letter \cite{N3LO}. The result presents a first milestone towards the missing piece and contains the combination of new and previously existing results\cite{N3Contr} to the soft-virtual (SV) approximation. 

%a milestone in the direction of obtaining this missing contributions was achieved and made public in the letter \cite{N3LO}. The letter presents the first calculation of a hadron collider observable at N${}^3$LO in QCD and contains the combination of new and previously existing results\cite{N3Contr} to the soft-virtual (SV) approximation, 
%a first milestone towards obtaining this missing contributions was achieved and made public in the letter \cite{N3LO},

%The improvement of the Higgs boson production cross-section for Hadron colliders is a large field of continuously improving research. 
%Currently, the largest perturbative uncertainty on the cross-section originates from the missing next-to-next-to-next-to-leading order ($N3LO$) QCD contributions to the gluon fusion production channel. Recently, a considerable advance in the direction of obtaining this missing contributions was made public in the letter \cite{N3LO}. The letter presents the combination of new and previously existing contributions to the first term of the threshold expansion of the $N3LO$ gluon fusion cross-section in the infinite top quark mass limit. 

%A threshold expansion has the intrinsic limitation of being accurate only up to the order the expansion was actually performed. 
To arrive at a reliable phenomenological prediction it is highly important to understand the limitation of the threshold expansion and draw conclusions about the necessity for further improvement via the calculation of sub-leading terms or even the full, unexpanded cross-section.
In this proceedings we study this uncertainty in the case of the gluon fusion Higgs production cross-section at N${}^3$LO. We consider lower orders in perturbative QCD to study the convergence behaviour of the expansion for the Higgs cross-section and inspect the impact of the ambiguity due to the truncation of the threshold expansion. Furthermore, we demonstrate that the ambiguity for the SV approximation at N${}^3$LO is large.

\section{Threshold Expansion for the Higgs boson cross-section}
%The Higgs production cross-section at the LHC takes the form
%\begin{equation}
%\label{eq:sigma}
%\sigma = \sum_{ij} \int dx_1\, dx_2\, f_i(x_1)\,f_j(x_2)\, \hat{\sigma}_{ij}(m_H^2,x_1\,x_2\,s)\,,
%\end{equation} 
%where  $\hat{\sigma}_{ij}$ are the partonic cross-sections for producing a Higgs boson from partons $i$ and $j$, $f_i(x_1)$ and $f_j(x_2)$ are the corresponding parton distribution functions, and $m_H^2$ and $s$ denote the mass of the Higgs boson and the hadronic centre-of-mass energy, respectively. 
%The largest contribution to the partonic cross-section is given by the gluon fusion production mode creating a Higgs boson via a top quark loop that is formed via the interaction of two initial state gluons. 
%The relatively light mass of the Higgs boson allows for the calculation of this process in perturbative QCD in the infinite top-quark mass limit.
%Currently Higgs cross-section is known in this fixed order QCD approximation up to next-to-next-to-leading order (NNLO) \cite{nnlo} and many additional corrections are available (see ref.\cite{ihixs} for a comprehensive summary). 

The probability distribution of a gluon occurring in a proton is steeply falling with its energy 
%This implies a larger probability of relatively small partonic center of mass energies for processes with gluonic initial state at hadron colliders.
%Consequently, 
and suggests the possibility of performing a fast converging threshold expansion of the gluon fusion Higgs cross-section.
% around the kinematic limit of producing a Higgs boson at rest and in addition only soft radiation. 
Already at NNLO a threshold expansion was performed\cite{nnlosoft} and was shown to be rapidly converging towards the full result\cite{nnlo2}.

%%%%%%%%%%%%%%%%%
Here we study the strong coupling expansion of the heavy top effective theory. 
In this note we are interested in the effect complementing existing lower order calculations with a threshold expansion at N${}^n$LO. 
The threshold approximations and expansions which we will discuss 
%in this note at N${}^n$LO 
will always contain the full (non-expanded) dependence on terms which enter the result at lower orders in the strong coupling expansion. We will also include the full N${}^n$LO dependence on 
renormalisation and factorisation scales as well as the full dependence on those N${}^n$LO corrections which are generated 
from higher order corrections to the Wilson coefficient. 
%%%%%%%%%%%%%%%%%

Parametrising the expansion with the variable $z=\frac{m_H^2}{x_1 x_2 s}$ leads to a series of the partonic cross-section in $(1-z)$.
\begin{equation}
\left[\hat{\sigma}_{ij}(s,z)\right]_{threshold}=\sigma^{SV}_{ij}+(1-z)^0 \sigma_{ij}^{(0)}+(1-z)\sigma_{ij} ^{(1)}+\dots \,.
\end{equation}
%where $\sigma^{SV}_{ij}$ is the first term, the so-called soft-virtual term, of the expansion.
%In the letter \cite{N3LO} the soft-virtual (SV) term, the first term of the a threshold expansion, of the $N3LO$ contribution was made public. This was the first time a hadron collider observable was calculated at this order in perturbative QCD. The threshold expansion allows to perform a perturbative Laurent series around the kinematic limit of producing a Higgs boson at rest and in addition only soft radiation. The expansion around this kinematic limit can be parametrised using the expansion parameter $(1-z)=(1-\frac{m_H^2}{x_1 x_2 s})$. 
If a series expansion is truncated at a given finite order an unavoidable ambiguity is introduced due to missing higher order terms. To study the impact of truncating the threshold expansion of the Higgs boson cross-section we spuriously insert a function $g(z)$ satisfying $\lim\limits_{z\rightarrow1}g(z)\rightarrow 1$ into eq.~\ref{eq:sigma} such that
\begin{equation}
\sigma=\sum\limits_{i,j} \int dx_1\, dx_2\, \left[ f_i(x_1)\,f_j(x_2) z g(z) \right]  \left[ \frac{\hat{\sigma}_{ij}(s,z)}{z g(z)} \right]_{threshold}.
\end{equation}
For all choices of $g(z)$ the expansion truncated at $\mathcal{O}\left((1-z)^{n}\right)$ thus leads to formally equivalent results up to $\mathcal{O}\left((1-z)^{n+1}\right)$.

\begin{figure}[h!]
\vspace{0.5cm}
\centering
\begin{subfigure}[Full result up to NNLO and NNLO SV term with different choices for $g(z)$ as function of $\mu$  ]{
\label{fig:NNLOgs}
\includegraphics[scale=0.29]{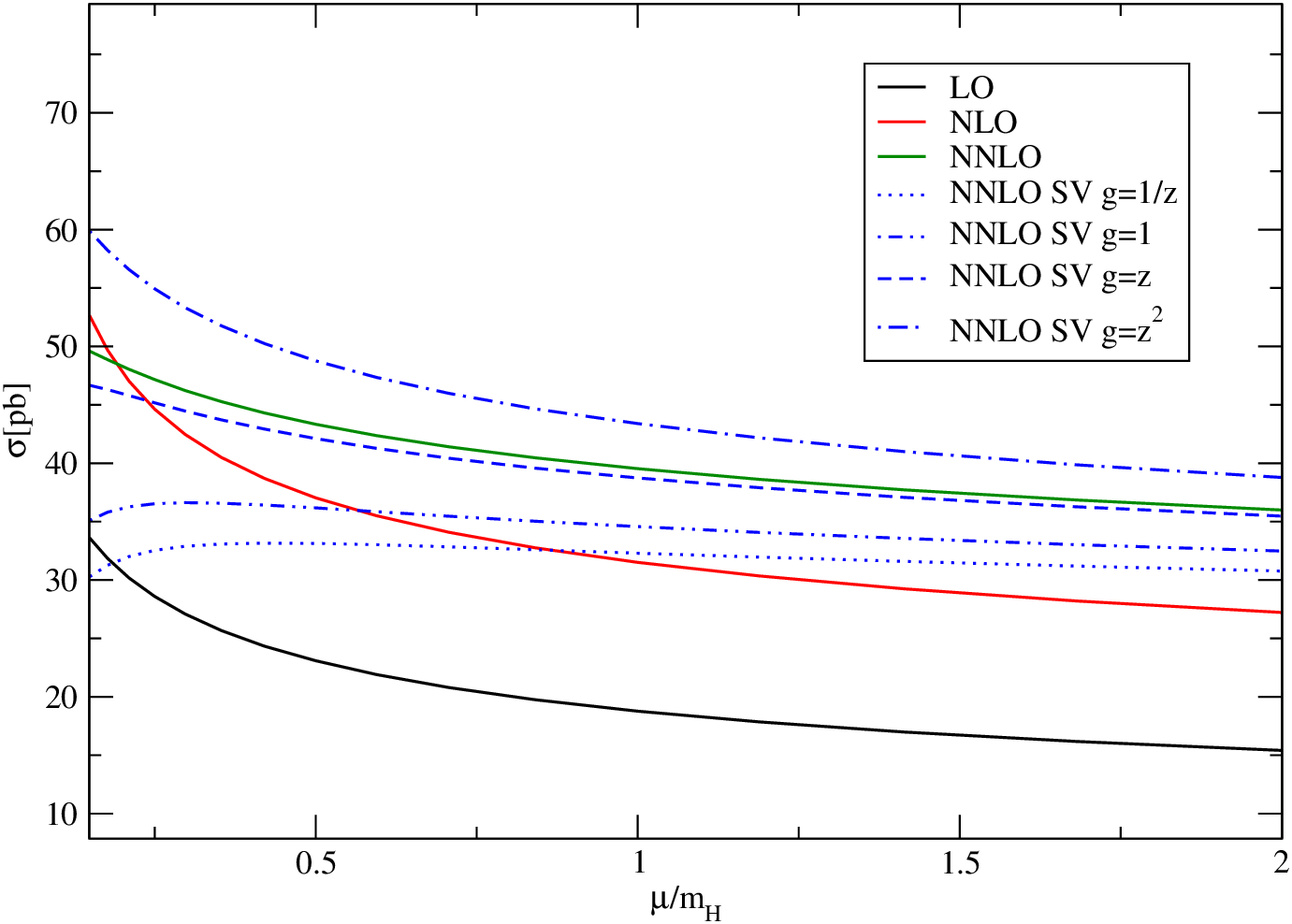}
}
\end{subfigure}
\begin{subfigure}[Expansion at NNLO truncated at different $\lambda=1-z$ for g(z)=1]{
\label{fig:NNLOgconv}
\includegraphics[scale=0.29]{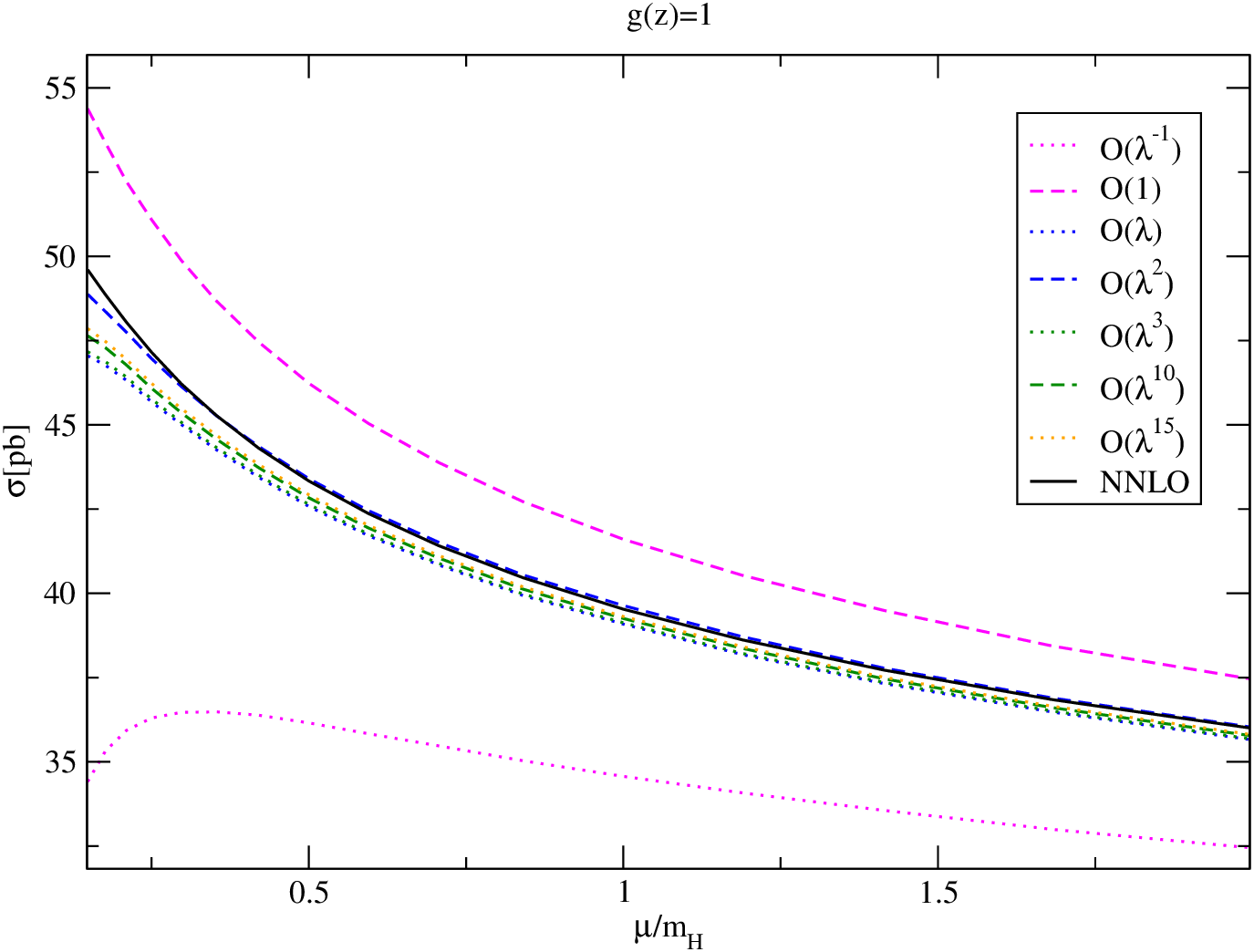}
}
\end{subfigure}
\caption{Threshold approximation for the Higgs boson cross-section at 13 TeV at the LHC}
\end{figure}

In Fig.\ref{fig:NNLOgs} we show the Higgs boson cross-section up to NNLO and the NNLO term including only the SV term as a function of the renormalisation and factorisation scale $\mu=\mu_R=\mu_F$. The different lines for the NNLO SV contribution correspond to different choices $g(z)=\{1,z,z^2,\frac{1}{z}\}$, respectively.

We note that the variation among the different choices is sizeable and suggests a large impact of sub-leading terms at NNLO. Of the selected choices $g(z)=z$ represents the closest approximation to the full result at NNLO. Analysing the first sub-leading term of the threshold expansion we find that this choice correctly reproduces the coefficient of the $\log^n(1-z)$, where $n$ is the largest appearing power. 
We found similar behaviour when analysing the NLO term.

In Fig.\ref{fig:NNLOgconv} we show the NNLO threshold approximation up to various orders in the expansion for the choice $g(z)=1$ as a function of $\mu$. One can clearly see that the first and second term suffer from a large discrepancy compared to the full result. Furthermore, we observe that the quality of the expansion is rather independent of the chosen scale $\mu$. We found similar behaviour for other choices of $g(z)$.

\begin{figure}[h!]
\vspace{0.5cm}
\centering
\begin{subfigure}[NLO]{
\includegraphics[scale=0.29]{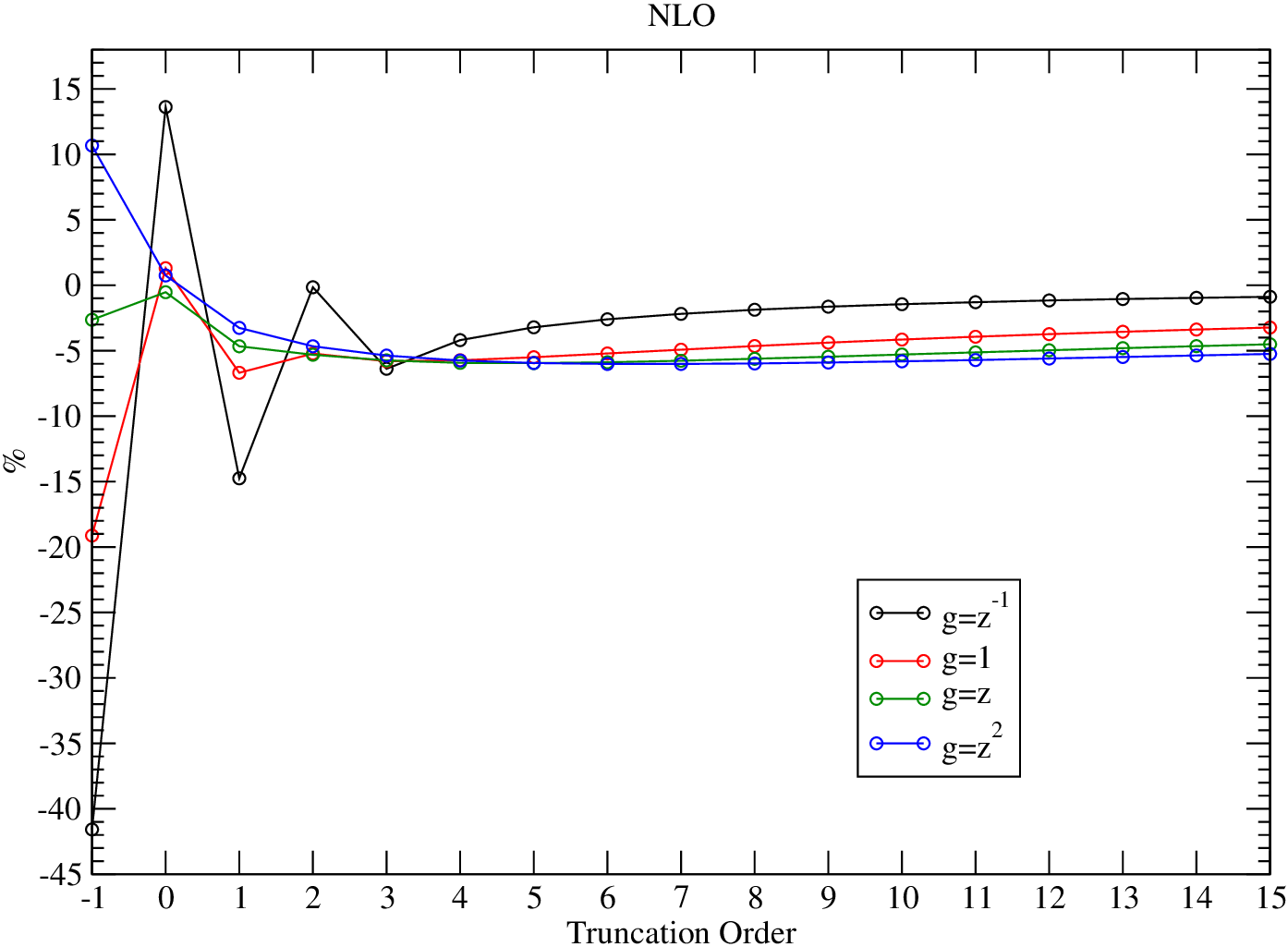}
}
\end{subfigure}
\begin{subfigure}[NNLO ]{
\includegraphics[scale=0.29]{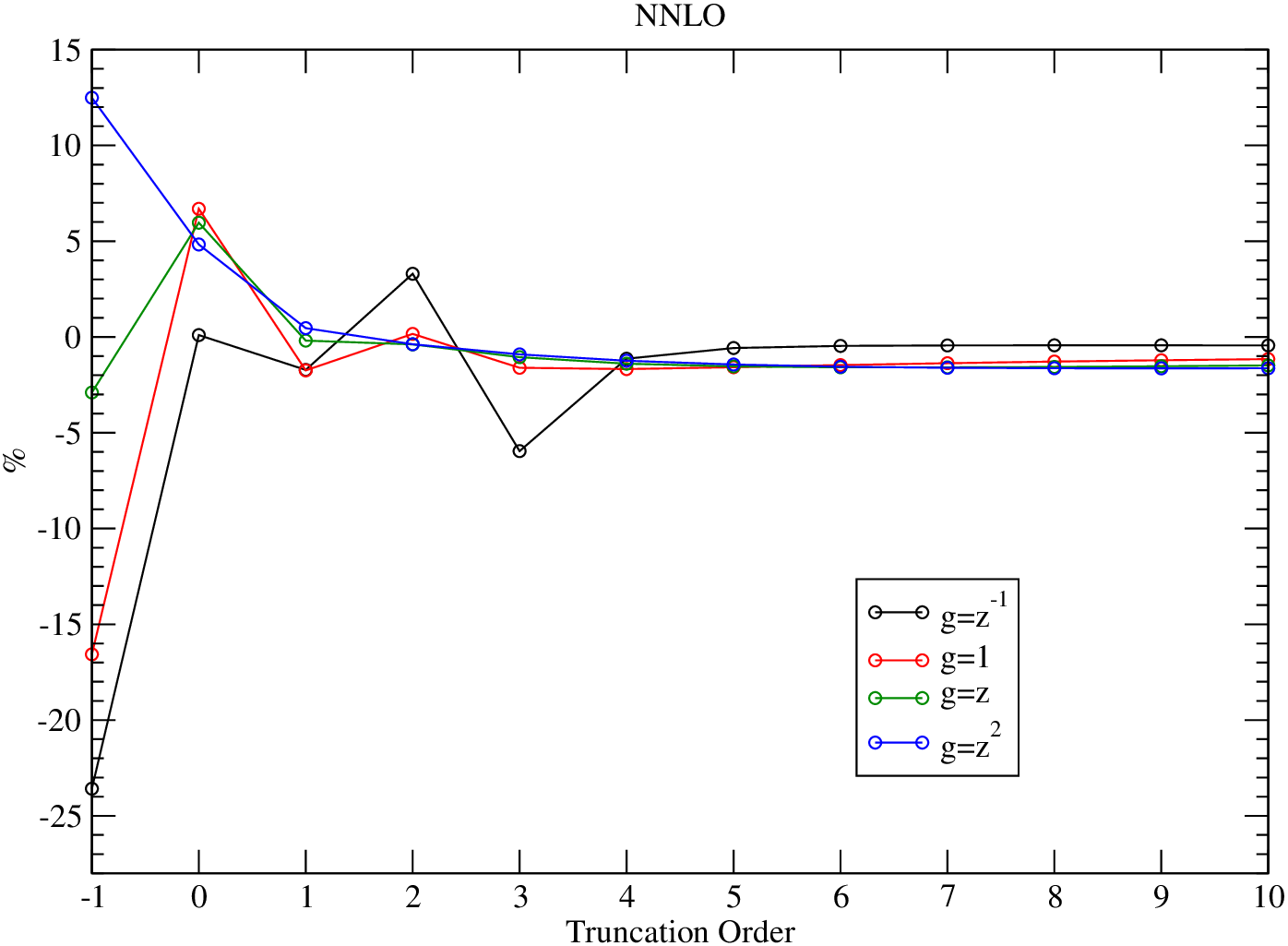}
}
\end{subfigure}
\label{fig:lowerexpansion}
\caption{Percent difference of the threshold expansion to the full Higgs boson cross-section at NLO and NNLO at 13 TeV at the LHC evaluated at $\mu=\frac{m_H}{2}$ as a function of the order where the series is truncated. Different lines correspond to different choices for the function g(z).}
\end{figure}

%As the SV term of the threshold expansion does not seem to present a reliable estimate for the full series, we study the impact of different choices for the function $g(z)$ as we add further terms in the expansion. 
In Fig.\ref{fig:lowerexpansion} we present the Higgs boson cross-section at NLO and NNLO evaluated at $\mu=\frac{m_H}{2}$ for the same choices of $g(z)$ as above as a function of the order where the threshold expansion is truncated. 
We observe sizeable changes of the prediction comparing the first and second term for all our choices of $g$. However the convergence pattern
observed for different choices of $g$ are rather different. Indeed while for lower orders in the expansion, $g=1/z$ is the worst choice, it becomes the best when up to $5$ or more terms are included.
This is particularly true at NLO, where the other choices only reproduce the full result within about $5\%$. 
The same effect, though much reduced in size, is also observed at NNLO and is directly related to the fact that the choices 
$g=z^n$, for $n\ge 0$, introduce a further damping of the gluon luminosity away from threshold, which is compensated by introducing a factor $1/z^{n+1}$ into the partonic cross section. This factor which increases the sensitivity to the high energy regime is subsequently expanded around the threshold and its effect is therefore lost completely in the SV and  still to some extent when further terms in the threshold expansion are taken into account. At NNLO it appears that the effect of this factor is much smaller than at NLO and we may expect this to hold also at N${}^3$LO.

%The relative change of the cross-section from the first to the second term is of the same size as the estimate for the uncertainty due to the truncation of the strong-coupling expansion via scale-variation at one order less. Combining only the SV approximation with existing lower orders consequently decreases the quality of the prediction. 
%We note that the ambiguity due to missing terms of the threshold expansion is rapidly decreasing. At $\mathcal{O}((1-z)^{5})$ any of the choices for $g(z)$ reproduce a reliable estimate of the full cross-section. The fastest converging behaviour is displayed by $g(z)=z$. 

\begin{figure}[h!]
\centering
\includegraphics[height=70mm]{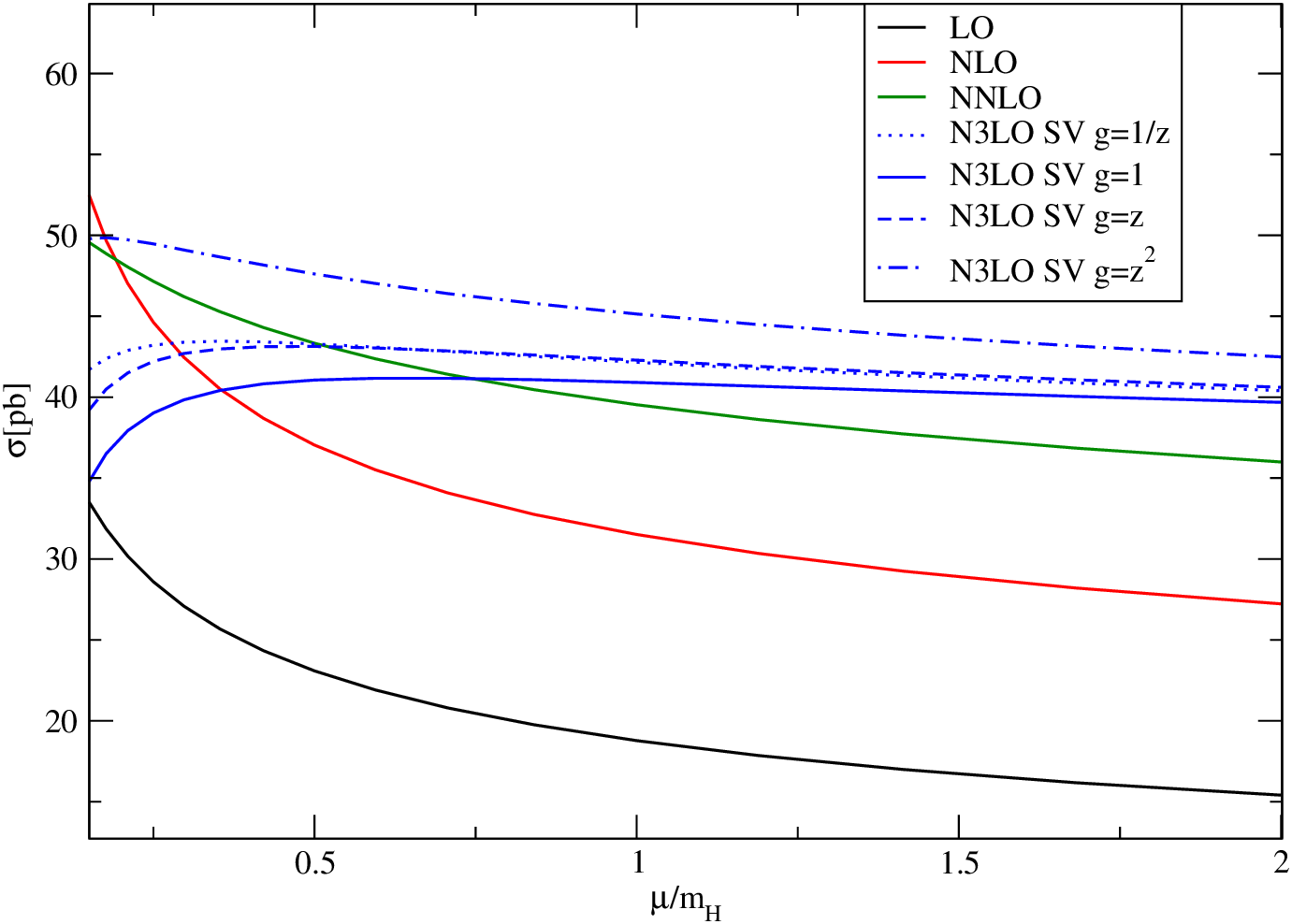}
\caption{The gluon-fusion cross-section at 13 TeV at the LHC as a function of $\mu=\mu_R=\mu_F$ up to LO (black), NLO (red), NNLO (green) and  soft-virtual N${}^3$LO (blue). The N${}^3$LO SV approximation is modified with different functions g(z).}
\label{fig:N3Plot}
\end{figure}

In Fig.\ref{fig:N3Plot} we present the SV Higgs cross-section at N${}^3$LO for the same choices of $g(z)$ as a function of $\mu$. Again the various choices lead to drastically different predictions for the Higgs boson cross-section. We observe that the hierarchy of the lines is changed compared to lower orders. The choices $g(z)=\{\frac{1}{z},1,z\}$ at $\frac{m_H}{2}$ are in agreement with the scale uncertainty at NNLO. The large spread of the different choices suggests the possibility for large corrections due to sub-leading terms at N${}^3$LO. Given the experience at lower orders we expect that only a few sub-leading terms in the threshold expansion are required to obtain a significant improvement to an approximation of the N${}^3$LO cross-section and consequently to the predictions for LHC observables.

\section{Conclusion}
The rapidly increasing experimental precision of Higgs cross-section measurements raises an urgent demand for the improvement of the theoretical prediction for the inclusive Higgs boson cross-section at the LHC. With the recent publication of the first term in the threshold expansion of the N${}^3$LO gluon-fusion QCD cross-section an important step in this direction was taken. In this proceedings we have analysed the quality of the threshold expansion. We find that the expansion is converging fast at lower orders in QCD perturbation theory and expect to find similar behaviour at N${}^3$LO. We studied the uncertainty introduced due to the truncation of the threshold expansion at NLO, NNLO and N${}^3$LO and conclude that at least several terms in the expansion are necessary in order to infer reliable predictions for LHC measurements and improve upon the current status. We conclude that the calculation of further terms in the threshold expansion and even the full Higgs boson cross-section at N${}^3$LO is highly desirable.
% and a topic of on-going research.

\section{Acknowledgements}
BM would like to thank the organisers of the $49^{th}$ rencontres de Moriond for the opportunity to present this result. 
We would like to thank Babis Anastasiou for helpful comments.
The research was supported by the European Commission through the ERC grant ``LHCTheory" (291377) and ERC grant "IterQCD" as well as the SNF contract 200021\_143781.

\end{document}